\documentclass[prl,twocolumn,showpacs]{revtex4}
\usepackage{epsfig}

\begin{document}

\title{Time-step targetting methods for real-time dynamics using DMRG}
\author{Adrian E. Feiguin and Steven R.~White}

\affiliation{Department of Physics and Astronomy\\
University of California, Irvine, CA 92697}
                                                                                
\date{\today}
                                                                                
\begin{abstract}
We present a time-step targetting scheme to simulate real-time dynamics 
efficiently
using the density matrix renormalization group (DMRG). The algorithm
works on ladders and systems with interactions beyond nearest neighbors, 
in contrast to existing Suzuki-Trotter based approaches.
\end{abstract}
                                                                                
\maketitle

Over the last ten years the density matrix renormalization group (DMRG) \cite{dmrg} 
has proven to be remarkably effective at 
calculating static, ground state properties of one-dimensional
strongly correlated systems. During this period there has also been substantial progress
made in calculating frequency dependent spectral functions\cite{spectral}.
However, the most significant progress in extending DMRG since its invention
has occurred in the last year or two.  Through a convergence
of quantum information and DMRG ideas and techniques, a number of new
approaches are being developed. The first of these are highly efficient
and accurate methods for real-time
evolution, allowing both the calculation of spectral functions via Fourier transforming,
and also novel time development studies of systems out of equilibrium. 

The key real-time methods thus far developed \cite{vidal-time,time,uri} 
rely on the Suzuki-Trotter (S-T) break-up of the
evolution operator. This approach has a number of important advantages: 
it is surprisingly simple and easy to implement in an existing ground
state DMRG program; the time evolution is very stable and the only source
of non-unitarity is the truncation error; and the number of density matrix
eigenstates needed for a given truncation error is minimal. It also has
two notable weaknesses: it has an error proportional to the time step $\tau$
squared, and, more importantly, it is limited to systems with nearest neighbor 
interactions on a single chain. As we show here, the accuracy can be
improved using higher order expansions. The nearest-neighbor/single
chain limitation is more problematic. In the case of narrow ladders with
nearest-neighbor interactions, one
can avoid the problem by lumping all sites in a rung into a single supersite.
Unfortunately, this approach becomes very inefficient for wider ladders, and is
not applicable to general long-range interaction terms.

In this paper we propose a new time evolution scheme which produces a basis
which targets the states needed to represent one time step. Once this basis
is complete enough, the time step is taken and the algorithm proceeds to the
next time step. This targetting is intermediate to previous approaches:
the Trotter methods target precisely one instant in time at any DMRG step,
while Luo, Xiang, and Wang's approach\cite{comment} targetted the entire
range of time to be studied. Targetting a wider range of time requires more
density matrix eigenstates be kept, slowing the calculation.
By targetting only a small interval of time, our
approach is nearly as efficient as the Trotter methods. In exchange for
the small loss of efficiency, we gain the ability to treat longer range
interactions, ladder systems, and narrow two-dimensional strips. In addition,
the accuracy is much improved over the lowest order Trotter method.

We want to find the solution to the time-dependent Schr\"odinger equation:
\begin{equation}
i\frac{d}{dt}|\psi(t)\rangle = \left(H(t) - E_0\right) |\psi(t)\rangle,
\end{equation}
where the ground state energy $E_0$ is introduced to reduce the amplitude
of the oscillations by making the diagonal elements of $H$ smaller
\cite{marston}.
We use a time dependent Hamiltonian to include the case where 
a time-dependent perturbation $V(t)$ is added to the time independent
Hamiltonian $H_0$.
The initial state $|\psi(t=0)\rangle$ is typically the ground state, 
or the ground state acted upon by an operator, but other possibilities
are also interesting.

When the wavefunction of the system evolves in time, its density matrix
samples a region of the Hilbert space that changes continuously. 
The DMRG basis is built to represent the states that are put into
the density matrix
\[
\rho =\sum_{t}w_{t}|\psi _{t}\rangle \langle \psi _{t}|,
\label{rho}
\]
where the target states $|\psi _{t}\rangle $ are weighted with a factor $%
w_{t}$, with $\sum_{t}w_{t}=1$. 
Typically, some sweeps are needed to build self-consistency between the
target states and the basis produced by the density matrix. A notable exception
to this need for self-consistency are the Trotter-based time evolution methods:
the bond time-evolution operator is represented exactly in the current
basis, and so the pretruncation density matrix is exact. Thus, the truncation
error is an exact measure of the error in the basis produced at that step. 
In our time-step targetting approach, this ideal behavior is lost, and
a sweep or two is needed to produce a good basis for the time-step.

How do we produce a density matrix representing the wavefunction over an interval
of time?  Luo, Xiang and Wang \cite{comment} (see also Ref. \cite{schmitteckert}) 
suggested targetting the wavefunction at a sequence of times spanning the interval,
$\psi (t=0)$, $\psi (t=\tau )$, $\psi (t=2\tau )$, $\ldots$, $\psi(t=n\tau)$, 
simultaneously. 
We argue that this choice is very close to ideal. Suppose that our basis
includes $\psi (k\tau )$ and $\psi ((k+1)\tau )$. Then the basis includes any
linear combination of these states, so that one could imagine using an interpolation
formula to determine coefficients $a$ and $b$ to approximate the wavefunction
at any time between $k\tau$ and $(k+1)\tau$ as
$\psi(t) \approx a \psi (k\tau ) + b \psi ((k+1)\tau )$. This suggests that
the error in the basis is at worst $\tau^2$. If the basis includes
more than two time points, one could imagine using higher order interpolations, e.g.
splines, putting a tighter bound on the error in the basis. 
The key point is that we do not actually perform these interpolations;
the basis is automatically good enough to allow whatever interpolation is
most accurate given the set of time points. This suggests that the error in
the basis varies as $\tau^{n+1}$.

If $\tau$ is small enough and $n$ big enough, and enough self-consistency
sweeps are made, the error in the basis is given by the truncation error.
This is the ideal situation for a DMRG calculation.
Since this truncation error is often miniscule we shall say that an approximate
algorithm is ``quasiexact'' when the error 
is strictly controlled by the DMRG truncation error $\epsilon$ (with
some properties proportional to $\epsilon$ and other to $\epsilon^{1/2}$).
For example, the infinite system method applied to a finite system is {\it not}
quasiexact, even though the error goes to zero as the number of states kept
$m$ increases. If enough sweeps are taken, and absent any ``sticking'' problems
with metastable ground states, the finite system ground state DMRG method is
quasiexact.
Non quasiexact algorithms seem to be the source of most DMRG ``mistakes''.
The procedure below is nearly quasiexact: it has a small separate time step
error, perhaps of order $\tau^4$, in addition to the truncation
error.

Our procedure consists of taking a tentative time step at each DMRG step, the
purpose of which is to generate a good basis. The standard fourth order
Runge-Kutta (R-K) algorithm is very convenient for this purpose. This is defined
in terms of a set of four vectors:
\begin{eqnarray}
|k_1\rangle &=& \tau  \tilde{H}(t) |\psi(t)\rangle, \nonumber \\
|k_2\rangle &=& \tau  \tilde{H}(t+\tau/2) \left[ |\psi(t)\rangle + 1/2 |k_1\rangle \right], \nonumber \\
|k_3\rangle &=& \tau  \tilde{H}(t+\tau/2) \left[ |\psi(t)\rangle + 1/2 |k_2\rangle \right], \nonumber \\
|k_4\rangle &=& \tau  \tilde{H}(t+\tau) \left[ |\psi(t)\rangle + |k_3\rangle \right],
\label{kvectors}
\end{eqnarray}
where $\tilde{H}(t) = H(t)-E_0$.
The state at time $t+\tau$ is given by
\begin{equation}
|\psi(t+\tau)\rangle \approx \frac{1}{6} 
\left[ |k_1\rangle + 2|k_2\rangle + 2|k_3\rangle + |k_4\rangle \right] + O(\tau^5).
\end{equation}
\label{newpsi}
We choose to target the state at times
$t$, $t+\tau/3$, $t+2\tau/3$ and $t+\tau$.
The R-K vectors have been chosen to minimize the error in $|\psi(t+\tau)\rangle$, but
they can also be used to generate $|\psi\rangle$ at other times.
The states at times $t+\tau/3$ and $t+2\tau/3$ can be approximated, with an 
error $O(\tau^4)$, as
\begin{eqnarray}
|\psi(t+\tau/3)\rangle & \approx & |\psi(t)\rangle + \nonumber \\
& + & \frac{1}{162} \left[ 31|k_1\rangle +14 |k_2\rangle + 14 |k_3\rangle  - 5 |k_4\rangle \right], \nonumber \\
|\psi(t+2\tau/3)\rangle & \approx & |\psi(t)\rangle + \nonumber \\
& + & \frac{1}{81} \left[ 16|k_1\rangle + 20 |k_2\rangle + 20 |k_3\rangle  - 2 |k_4\rangle \right].
\label{targets}
\end{eqnarray}

\begin{figure}
\begin{center}
\epsfig{file=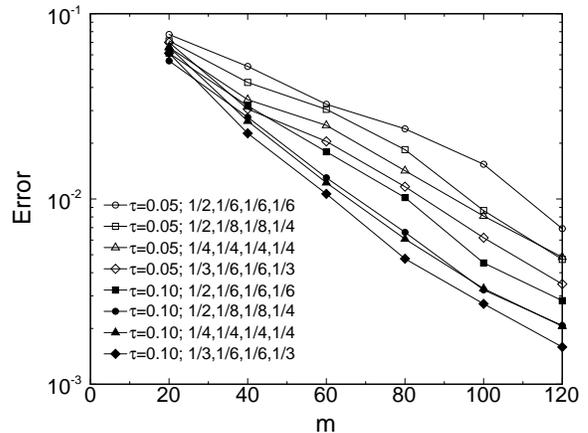,angle=-90,width=8cm}
\caption{
Error $E(t=8)$ for the Haldane chain ($L=32$), according to Eq. (\ref{error}), as a function of the number of states kept $m$.
We show results of simulations using the Runge-Kutta algorithm, for different time steps and distributions of weights. 
}
\label{fig1}
\end{center}
\end{figure}

In practice we proceed as follows: each half-sweep corresponds to one time 
step. 
At each step of the half-sweep, we calculate the R-K vectors 
(\ref{kvectors}), but without advancing in time. The density matrix is then
obtained by using the formula (\ref{rho}) with the target states
$|\psi(t)\rangle$, $|\psi(t+\tau/3)\rangle$, $|\psi(t+2\tau/3)\rangle$, and
$|\psi(t+\tau)\rangle$. 
Advancing in time is done on the last step of a half-sweep. However,
we may choose to advance in time only every other half-sweep, or only
after several half-sweeps, in order to make sure the basis adequately
represents the time-step. Our tests show that one half-sweep
is adequate and most efficient for the systems studied here.
The method used to advance in time in the last step need not be the R-K
method used in the previous tentative steps. In fact, the computation time involved
in the last step of a sweep is typically miniscule, so a more accurate
procedure is warranted.
A simple way which keeps the time-integration errors much smaller than the
basis errors is by performing 10 R-K iterations with step $\tau/10$. We usually
use this method. Alternatively, one can evolve using the 
exponential of the the Hamiltonian in the Lanczos tridiagonal representation,
which is exactly unitary. However, the truncation to a finite number of
density matrix eigenstates introduces nonunitarity anyway, so the Lanczos
procedure has no special advantage. In practice, we find comparable overall
accuracy in the two methods \cite{lanczos}.

\begin{figure}
\begin{center}
\epsfig{file=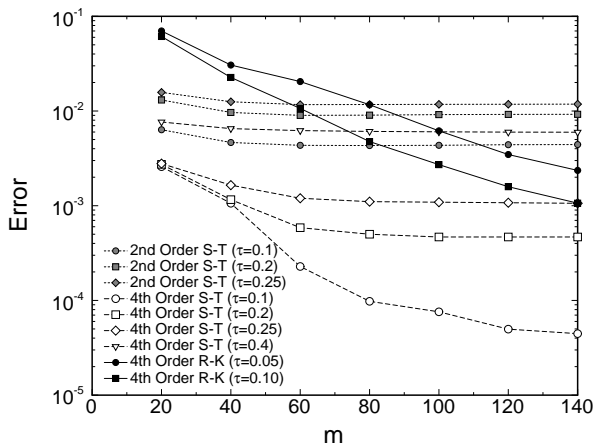,angle=-90,width=8cm}
\caption{
Same as in Fig. \ref{fig1}, using 1st order Suzuki-Trotter break-up (gray symbols), 
4th order Suzuki-Trotter (empty symbols), and
4th order Runge-Kutta (filled symbols).
}
\label{fig2}
\end{center}
\end{figure}

To test the method we first studied the $S=1$ Heisenberg chain.
Since it is a single chain system, the Suzuki-Trotter methods are also 
applicable. 
In addition to our new method, we have used both the traditional 
1st order Suzuki-Trotter decomposition \cite{time} and the 
4th order Forest-Ruth break-up \cite{forest-ruth}. In order to compare the
results, we calculated the error as
\begin{equation}
{E}(t) = \sqrt{\frac{1}{L}\sum_{x=1}^L \left(S^z(x,t)-S^z_{Exact}(x,t)\right)^2},
\label{error}
\end{equation}
where $S^z_{Exact}$ is obtained using 4th order Suzuki-Trotter with $m=200$ and
 $\tau=0.02$, which  
keeps the truncation error under $10^{-12}$. 

The target states (\ref{targets}) can be weighted equally, or unequally. We 
have performed several test runs with different distributions of weights. 
In Fig. \ref{fig1} we show the error (\ref{error}) at time $t=8$ as 
function of the number of states kept $m$, for various weightings. 
The best weighting we have found is $w_1=w_3=1/3$, $w_2=w_3=1/6$. The
calculations described below, unless otherwise noted, use this choice of weights.

\begin{figure}
\begin{center}
\epsfig{file=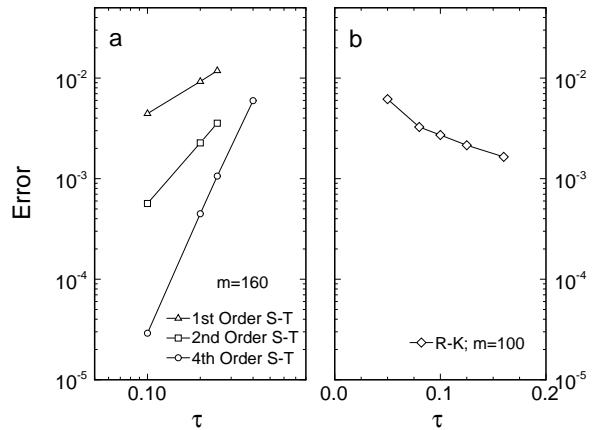,angle=-90,width=8cm}
\caption{
Error $E(t=8)$ for the Haldane chain for different time steps $\tau$: 
a) 1st, 2nd, and 4th order Suzuki-Trotter break-ups and $m=160$; b) Runge-Kutta
and $m=100$.
}
\label{fig3}
\end{center}
\end{figure}

In Fig. \ref{fig2} we compare results by using our method and Suzuki-Trotter evolution. 
The Suzuki-Trotter simulations 
converge when the error reaches a plateau and remains constant with
increasing number of states $m$. This occurs generally for a relatively 
small $m$, after which the accuracy of the simulation is completely 
controlled by the Trotter error, and not by the truncation error. 
In Fig. \ref{fig3}a we verify that the quantity $E$ is proportional 
to $\tau$, $\tau^2$, and $\tau^4$ for the three 
Suzuki-Trotter break-ups considered \cite{forest-ruth}. 
In the R-K simulations, convergence is slower with the number of
states $m$ because we need the basis to be optimized for 4 states at 
slightly different times. 
The accuracy improves steadily with the size of the basis, 
and also with the size of the time-step $\tau$, as it can be seen 
in Fig. \ref{fig3}b, although the method
breaks down for time-steps larger than $\tau \simeq 0.25$. 
These results may look counter intuitive, since the R-K error is expected to be 
proportional to  $\tau^4$. The reason for this behaviour is 
that smaller time-steps require more iterations, with a consequent 
accumulation of error due to the truncation. Therefore, unlike the S-T 
case, the simulation is 
now dominated by the truncation error, which can be reduced by increasing 
the size of the DMRG basis.  

For typical accuracies on a 1D chain, the R-K algorithm is 
numerically costlier than the S-T counterparts.
Measuring the the CPU time required to reach a time $t$ in the simulation 
we find, for instance, that in order to obtain 
an error of the order of $10^{-3}$ at $t=8$ we could use 1st order S-T with 
$\tau=0.016$ and $m=40$ (CPU time for 1000 half-sweeps: 34 minutes), 4th order S-T with 
$\tau=0.25$ and $m=40$ (CPU time for 224 half-sweeps: 7 minutes), or R-K 
with $\tau=0.10$ and $m=140$ (CPU time: 104 minutes). 
The R-K method requires fewer sweeps to reach a specified time. However, it also requires more states to be kept, leading to a substantially lower computation time. 

\begin{figure}
\begin{center}
\epsfig{file=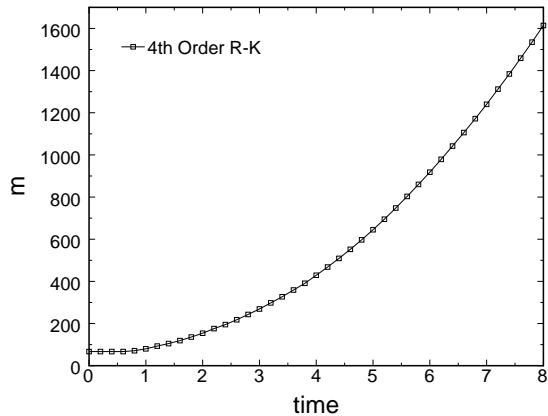,angle=-90,width=8cm}
\caption{
Number of states required to keep a truncation error of $10^{-8}$, as a function of time. The results correspond to a R-K simulation of a Haldane chain with $L=32$.
}
\label{fig4}
\end{center}
\end{figure}

In Fig. \ref{fig4} we show how the number of states $m$ required to keep a fixed, very small truncation error of $10^{-8}$ grows with time. 
This rapid growth in $m$ for a fixed accuracy is not surprising. At $t=0$, an operator is applied to the ground state, creating $|\psi(0)\rangle$. For small $t$, $|\psi(t)\rangle$ is still closely related to the ground state, and so requires a comparable number of states to represent it. For larger $t$, $|\psi(t)\rangle$ becomes more complicated as each excited eigenstate evolves with a different frequency and becomes independent of the others.

\begin{figure}
\begin{center}
\epsfig{file=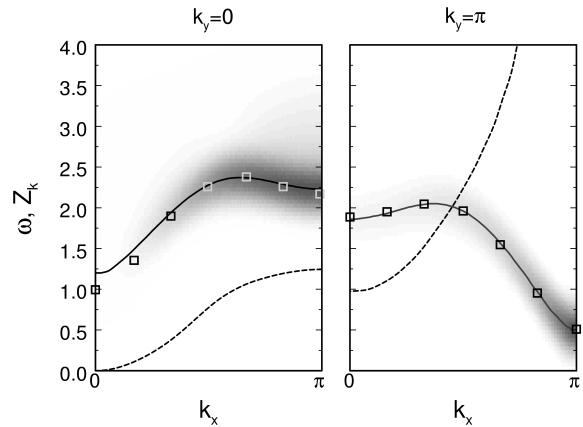,angle=-90,width=8cm}
\caption{
Structure factor $A(\mathbf{k},\omega)$ for the Heisenberg ladder using
4th order Runge-Kutta, $m=256$ states, and time-step $\tau=0.1$. 
The solid line is centered atthe quasiparticle peak.
The tones of gray are proportional to the quasiparticle weight (dashed curve). 
The symbols are Lanczos results for $L=12$ from Ref.\cite{riera}.
}
\label{fig5}
\end{center}
\end{figure}

As an application of the R-K method we calculated the spin structure factor 
for a 
$2\times L$ Heisenberg ladder with spin $S=1/2$, 
$A(\mathbf{k},\omega)=-\frac{1}{\pi}\mathrm{Im}G(\mathbf{k},\omega)$,
obtained by Fourier transforming the time-dependent
spin-spin correlation function \cite{time}
\[
G(\mathbf{x},t)=\langle S^-(\mathbf{x},t)S^+(0,0) \rangle.
\]
In this case, besides targetting the four states at different times (\ref{targets}), 
we also need to target the ground state at $t=0$. 
We have used a weight $w_0=1/2$ for the ground state, and all the other weights
equal to $1/8$.  In Fig. \ref{fig5} we show the results for $L=32$
using a time step $\tau=0.1$ and $m=256$, which kept the truncation error 
under $10^{-7}$ for times up to $t=30$. The result for the spin-gap is 
$\Delta=0.506$, which should be compared to the very precise DMRG value 
$\Delta_{Exact}=0.50249$ in the thermodynamic limit \cite{gap-ladder}. 
We also show for comparison the exact diagonalization results for the 
singlet-triplet excitations for $L=12$ from Ref.\cite{riera}. 
A continuum of excitations can be observed above the 
magnon band for $k_y=0$. It becomes more difficult to resolve the
band for $k_y=0$ in the proximity of $k_x \rightarrow 0$ because the
quasiparticle weight tends to zero in this limit. This is not the case for 
$k_y = \pi$, where the band is well defined in the entire range of momenta.

To summarize, we have presented a new algorithm for simulating time 
evolution of quantum systems. We described how to tune the parameters in order 
to reach accuracies comparable to those obtained by using 
Susuki-Trotter based approaches, and demonstrated its application by 
calculating the excitation spectrum of the Heisenberg ladder. 
Unlike methods that rely on Suzuki-Trotter break-ups, our algorithm
can be applied to systems with arbitrary geometry, and 
interactions beyond first neighbors. Moreover, it can be easily generalized 
for studying more complex models with strong correlations. 

We acknowledge the support of the NSF under grant DMR03-11843.


\begin{thebibliography}{99}
\bibitem{lanczos} This is equivalent to using the first term of Eq.(5) in 
\cite{schmitteckert}. We have used 10 eigenvectors. 
Notice however that we cannot 
expect the basis to be optimized to represent all of them accurately.
\end{thebibliography}
\end{document}